\documentclass[12pt]{bmc_article}

\usepackage{cite} 
\usepackage{url}  
\usepackage{ifthen}  
\usepackage{multicol}   
\usepackage[utf8]{inputenc} 
\usepackage{graphicx} 

\newboolean{publ}

\newenvironment{bmcformat}{\begin{raggedright}\baselineskip20pt\sloppy\setboolean{publ}{false}}{\end{raggedright}\baselineskip20pt\sloppy}

\begin{document}
\begin{bmcformat}


\title{Biological Database of Images and Genomes: tools for community annotations linking
image and genomic information}
 

\author{Andrew T. Oberlin$^{1}$%
      \and
         Dominika A. Jurkovic$^2$%
      \and
         Mitchell F. Balish$^2$ %
       and 
         Iddo Friedberg$^{1,2,*}$%
         \email{Iddo Friedberg - i.friedberg@miamioh.edu}%
      }


\address{%
    \iid(1)Department of Computer Science and Software Engineering,%
        Miami University, Oxford, OH, USA\\
    \iid(2)Department of Microbiology, Miami University, Oxford, OH, USA
}%

\maketitle


\begin{abstract}

Genomic data and biomedical imaging data are undergoing exponential growth. However, our
understanding of the phenotype-genotype connection linking the two types of data is lagging
behind. While there are many types of software that enable the manipulation and analysis of
image data and genomic data as separate entities, there is no framework established for linking
the two. We present a generic set of software tools, BioDIG, that allows linking of image data
to genomic data. BioDIG tools can be applied to a wide range of research problems that require
linking images to genomes. BioDIG features the following: rapid construction of web-based
workbenches, community-based annotation, user management, and web-services.  By using BioDIG to
create websites, researchers and curators can rapidly annotate large number of images with
genomic information. Here we present the BioDIG software tools that include an image module, a
genome module and a user management module. We also introduce a BioDIG-based website, MyDIG,
which is being used to annotate images of Mycoplasma. \\

BioDIG website: http://biodig.org \\
BioDIG source code repository: http://github.com/idoerg/BioDIG \\
The MyDIG database: http://wan1.mbi.muohio.edu/dome

\end{abstract}
{\footnotesize 
This article has been submitted for publication in Database \copyright: 2012-2013 Iddo Friedberg 
Published by Oxford University Press on behalf of all authors. All rights reserved.}

\ifthenelse{\boolean{publ}}{\begin{multicols}{2}}{}


\section*{Introduction}

The large amount of  genomic data has led to a concurrent deluge for new and creative ways of
genomic data processing and visualization. These are manifest in many types of genome browsers
\cite{pmid22638571,pmid22222089,pmid22102585,pmid21450055,pmid19846439,pmid19605418,
pmid19570905,pmid15975227,pmid15123591,pmid12368253,pmid12045153,pmid21139605}
and in the large number genome annotation tools both manual and computational (See
reviews:\cite{Friedberg2006Automated, Rentzsch2009210}). 

However, genomic data are not the only biological data that require high throughput analysis and
annotation. Since long before the genomic revolution, biological image data have been produced steadily
from sources as diverse as drawings, medical radiology (MRI, X-ray), microscopy and regular
photography. Like genomic data, biological images also require analysis and
annotation that include feature detection, segmentation, and association with biological
function. Computer-aided and computational image analysis has been performed since computers
emerged in the 1940s, with major advances made as the acquisition, storage, analysis and
manipulation of images have become cheaper and more accessible. 

With the deluge of both genomic and biological image data, there is a need for infrastructure
capable of handling both types of data and at the same time associating them in a meaningful
fashion. There are many projects that catalog biological images in fields as diverse as zoology,
microbiology, neurobiology, developmental biology and cancer research, to name a few. However, to
date there has been no freely available software to enable the association of image features with
genomic features. With the large number of both image and genomic data, and with the
understanding that phenotypic data manifest in images is ultimately a result of genomic events
(features), there is a need to easily associate image data and genomic data. While databases
providing such information do exist \cite{pmid9045213,pmid23041053}, 
there are no general-purpose software tools that can
be used to construct such an association.

Here we present the Biological Database of Images and Genomes or BioDIG. BioDIG is a set of
tools to construct databases with genomic and image data and place them in semantic
association. BioDIG provides the following features: a genome uploader and browser, an
image uploader, browser, and tagging workbench, association of image features with genomic
features, user management that features circles of trust for image feature tagging, and web
services. We demonstrate an example web-site, MyDIG, which we have constructed using
BioDIG. BioDIG is released under an open source license, and is freely available. We call
upon the community of bioinformatics programmers to join in its development.

\section*{Methods}
\subsection*{General}

BioDIG is a web-based workbench that enables users to upload images, annotate parts of them with
semantic data and associate these tags with genomic features. BioDIG includes user management
with increasing circles of privileges. 
BioDIG is implemented using three modules: the Genome Module, the Image Module  and the Users
Module.  The Genome Module is used to upload, store and visualize genomes. The Image Module is
used to upload, store, annotate and visualize images. The User Module is used to manage the users
and their privileges in the system.

\subsection*{Genome Module}

The Genome Module implements the storage of genomic information that should be readily
accessible for both queries and visualization. As a foundation, we use the Generic Model
Organism Database, GMOD, database schema, Chado, which is implemented in PostgreSQL. This
database was designed to represent genomic information found on NCBI in the form of GFF
files. BioDIG is implemented as an extension of Chado adding links from the genes in the
feature table of Chado to the images in the image database extension. The features table of
Chado is used to store data about genes, gene components, mRNA and other available data in
the GFF format. BioDIG is implemented using PostgreSQL, however it uses a Django framework
(djangoproject.com) for web service and web client access to support community annotation.
For visualization of the genomic data, GMOD also offers a genome visualization tool,
GBrowse\cite{pmid12368253}, which is a Perl-based system that can link to a multitude of genomic
databases including Chado. However, when connecting to Chado, GBrowse runs much more slowly
than when connecting to other databases. Thus, we have chosen to duplicate the data in
SQLite3, which will be used for the rendering of GBrowse. These files break up the data
into separate databases so that each genome can be separated, increasing search time. The
SQLite3 database that is used in conjunction with GBrowse visualization is
Bio::DB::SeqFeature::Store, which stores the features as flat Perl objects, which eliminate
all of the table joins that Chado must go through in order to prepare the information for
visualization. Normally, one would be concerned with the maintainability of a system that
duplicates information. However, the genomic information in these databases is expected to
change only very rarely, leaving little risk that these sources will become
unsynchronized.  We expect that in this database schema, one would perform these
information updates. For instance, when a gene's function becomes known for an organism that is already
in the database a batch procedure will be performed, making minimal changes to the database
without breaking any foreign keys. Using this Genome Module, we are able to visualize and
store information about genes so that queries are optimized and the visualization workload
is not burdening the search system.

\subsection*{Image Module}

The Image Module's goal is to handle the creation and storage of images and image metadata. Here \textit{image
metadata} is taken to mean identifiable components of biological interest within the images. Organelles, cells,
tissues, organisms, and other biological objects in the images are all considered identifiable components. The image
module supports the classification of these components. We have implemented a web user interface that allows the
tagging of images by drawing on the image and describing the substructure using free text and/or an ontology.
Finally, once components have been identified, they can be linked to the associated genes via \textit{gene links}
mapping the phenotypic data in the images to the genomic data (Figure~1).

The Image Module's most important task deals with storing the image files in a fashion that makes
them highly accessible. First the picture table stores the path to the image on the file system.
The two options often used for image storage are to store the image in the database as a BLOB and
to store a path to the image in a media directory or the URL on a media server.
However, storing a file as a BLOB in a database has certain advantages over
storing it as a file such as lower read throughput for objects less than 1 MB in size for short
intervals and atomicity for overwriting \cite{Sears2007To}. However, storing an image as
a path to a file within a file system provides other advantages such as high throughput for files
greater than 1 MB and low fragmentation in the long term. 

Because we use the Django framework,
Python objects are linked to tables in the database. When one retrieves one of these objects, the
entirety of the data is made available to the user. Thus, storing the path can decrease search
times that may only use the picture table as a foreign key to link information. In order to
increase throughput even more, the images are kept in consistent format, meaning that all
pictures are encoded in PNG format to remain web compliant. The names of these images are also
changed to a unique key, which can be used for hashing. Once the database grows in size, these
hashed names will be used to limit the number of images per directory by taking two letter
prefixes as the names of the directory in which the image resides, a method suggested by
\cite{Sears2007To} for increasing server performance.

The rest of the schema for the images represents other image metadata such as its description, upload date, the user
who owns the photo and the image tags. The tags themselves are represented by a combination of three tables, tag
points, tags, and tag groups. The tag points are an ordered list of points that when drawn in an API form a polygon.
These points are held together by the tags, which hold the metadata such as the structure related to the tag in the
image. We also store information about tag groups for users to make there own tag groups. These groups will exist
unique a user until that user wishes to merge with another user's tag group on the same image. When these tags are
linked to the genes in the feature table of Chado via gene links, data querying is enabled (Figure~2).

\subsection*{User Management Module}

BioDIG enables collaborative editing of image meta-data, with the aim of improving image
annotations through rapid collaborative work by a large number of users.  At the same time, there
is a need to maintain high annotation quality, and  prevent vandalism. To achieve these goals,
BioDIG features a User Module that provides three user permission levels, with each level
including the permissions of the preceding levels. 

Level 1 users have the most basic set of permissions, 
being able to search published information on the
website and download the results of their search, including
genomic, image, image-tag and gene-link information, for further analysis. The majority of users are intended to
have level 1, read-only access to a BioDIG-based site. Level 2 users, or annotators, have
site-wide read privileges, and in addition can employ tagging user interfaces and pipelines
for publishing.  Level 2 users are a much smaller number, and assigning a user to Level 2 is
supposed to be based on a circle of trust.  Level 3 users, or administrators,  have
permissions to access the databases and edit information directly. These users will also be
in charge of monitoring requests for publication and merge requests for tag groups as the
result of the pipeline. Figure~3 illustrates the user permissions scheme.

In this community-annotation effort, crediting users with the publication of images can connect
labs performing similar research and allow the user to edit these photos in their own workbench.
Many websites attempt to have private and public layers; however, this implementation of the
process is aimed to allow for approved and unapproved information to maximize community
engagement. Once a tag is sent through the pipeline, it is available for all users in the public
domain to see.  However, if the administrators of BioDIG approve the tag, it is said to be of a
trustworthy source having no obvious errors. This idea is implemented to increase the amount of
quality information in the database, while not limiting the capabilities of the system.
Inclusion of the user who owns the image also allows the person to be involved in the tagging of
the picture by other people; the approval process of the tags and associated gene-links  is done
by trusted users only, helping maintain quality.  With the administrators monitoring this
submission process we will be able to avoid deadlocks and make the data more accessible.

\subsection*{Web services and data analysis}

Web services provide an opportunity for researchers to mine data from
BioDIG-enabled websites. Most of the search features of the web site itself are, or soon will be, available
via web services. This will allow for a command line interface (CLI) to
be implemented along with an Application Programming Interface (API) in common programming
languages for interacting with these services programmatically.  For instance, if someone wishes
to retrieve a phenotype caused by the expression of several orthologs in different species, a
client API could be used to query for tags related to a set of  genes across a list of species,
and retrieve the relevant images and image tags. The client could then access the
information for each of these tags through a predetermined object-oriented interface, allowing for
more dynamic searches than one would naturally make by visiting the website. The user could then
run metrics to analyze the image and the tag data in a program rather than just by hand to
increase the speed at which this data can be analyzed. 

For example to return information about image 52, a caller program can use the URL: 
http://wan1.mbi.muohio.edu/dome/images/getImageMetadata?imageKey=52 \\
Which will return a Python dictionary as a valid JSON object:

\begin{verbatim}
{  
    "description": "Mycoplasma pneumoniae", 
    "organisms": [{"common_name": "Mycoplasma pneumoniae M129", "organism_id": 7}], 
    "url": "/dome/media/pictures/85c4bc19fa0d481460bcb522349fba27.png", 
    "errorMessage": "Success", 
    "uploadDate": "2012-11-24 19:14:08", 
    "error": false, 
    "uploadedBy": "oberliat"
}
\end{verbatim}

\subsection*{Example Site: the MyDIG website}

We describe MyDIG, a BioDIG-derived website implemented to show mycoplasma images and
genomes.  Mycoplasmas are a clade of bacteria whose main characteristics are parasitism, the
lack of a cell wall, and a small, rapidly evolving genome \cite{pmid12663864}. The cellular
morphology of mycoplasmas is especially interesting, as the limited complexity of
mycoplasmal cellular structure, with small, wall-less, single membrane-bound cells, enables
researchers to focus on universal principles of cellular organization\cite{pmid16720287},
providing a platform for understanding more derived types of cells.

For these reasons, we created the Mycoplasma Database of Image and Genomes or MyDIG using
the BioDIG infrastructure. Phylogenetically-associated cellular morphology is cardinal in
the description of mycoplasmas. Morphological features in mycoplasmas are often associated
with known genomic features. We present a series of screenshots showing a typical use-case
scenario for the image-genomic association for MyDIG.  \textit{Mycoplasma pneumoniae} is a
significant human pathogen whose cell shape is distinctly polarized\cite{pmid18754792}. From
the cell body, which contains most of the cell volume, emanate both the attachment organelle
and the trailing filament\cite{pmid16720287}. They are both prosthecal
structures, continuous with the cytoplasm of the cell body. Although the trailing filament
is not well-understood, the attachment organelle  of \textit{M.  pneumoniae} and its close
relatives is essential for adherence to host cells\cite{pmid11325545}, which is necessary
for all growth in a non-laboratory environment, and also contains a poorly-characterized
motor that contributes to cell motility and cell division\cite{pmid14925, pmid17163973,
pmid20735775}. The composition and assembly of \textit{M. pneumoniae}'s attachment organelle
are well-studied, having been established through analysis of non-adherent
mutants\cite{pmid14763969}. Essential to the formation of the attachment organelle is a
cytoskeletal structure called the electron-dense core\cite{pmid4914084}, which has been
imaged extensively and carefully\cite{pmid16573687, pmid16875842}; despite this, the
specific locations of its protein components are only vaguely defined. In Figure~4, the
general locations  of these proteins as well as the transmembrane proteins involved in
adherence are indicated\cite{pmid12533484, pmid15466048, pmid19183275} on a scanning
electron micrograph of an \textit{M. pneumoniae} cell grown attached to a glass cover slip,
as previously described in\cite{pmid16804191} and imaged at high magnification.

The need for an image-genomic database is illustrated by the case of \textit{Mycoplasma
microti}, a mycoplasma associated with prairie voles. The original publication describing
this species includes a transmission electron micrograph of sections through several
\textit{M. microti} cells which reveal the variety of polarization idiosyncratic to the
\textit{Mycoplasma muris} phylogenetic cluster, of which \textit{M. microti} is a member. At
the time of its publication (1995), images showing very similar structures of other members
of this cluster had been published\cite{pmid1683419, pmid2665713}.  However, the authors of
the \textit{M. microti} study, perhaps unaware of these images, described the morphology in
terms that emphasize pleomorphy and imply the absence of polarization\cite{pmid8746521,
pmid11321086}. This flawed description led to \textit{M.  microti} being described
subsequently as non-motile, since absence of polarity is associated with absence of
motility\cite{Brown2011Bergey}. However, we recently determined that like other species of the \textit{M. muris}
cluster that exhibit polarization, \textit{M.  microti} exhibits robust gliding
motility\cite{pmid22447904}, (S.L. Distelhorst and M.F. Balish, unpublished data). Had a
database of mycoplasma morphological features been available to the authors, this error
could have been prevented.

\section*{Conclusions}

We have described a new software package for constructing websites linking biological image and
genomic data. When designing BioDIG, we did so with generic implementation in mind, so that
anyone who associates genomic data with image data may benefit from its adoption. BioDIG is
suitable for uses as diverse as cancer pathology, zoology, plant biology, cell biology and
microbiology, to name a few. The web services feature is especially useful for linking to other
web resources that can capitalize on BioDIG-enabled sites. 

Being a collection of modules, rather than a single application,  BioDIG can be applied to a wide
range of uses. Those can
range from microbiologists researching certain taxa (as the mollicute-oriented MyDIG is
intended, see example above) to zoologists or plant scientists interested in whole-organism
morphologic features and their genomic associations. The biomedical research community, both
applied and basic, can also benefit from BioDIG. Image databases of cellular and tissue
morphology, radiological images and pathology can all benefit
from incorporating BioDIG if links to genomic information are useful. BioDIG-based websites can
also be used as a teaching resource, in application ranging from primary secondary to
post-secondary and higher education. We call upon the community of biology software developers to
join us in the development of BioDIG, which is distributed under an open source license.  We
encourage life science researchers to contact us regarding applications of BioDIG suitable for
their needs.

Image feature tagging currently uses freeform drawing, and can be tagged using free text.
However, for certain types of applications it may be better to provide the annotator with a
controlled vocabulary or an ontology to make annotations standard and computationally tractable.
In the future we intend to provide a mechanism for doing so. Other features that require
implementation are searches by taxon, image tags, gene names, gene annotations, and boolean
combinations of all of the above.

\section*{Authors contributions}

 ATO, MFB and IF conceived the study. ATO and IF designed BioDIG. ATO designed the code and database schema, and
 wrote the code for BioDIG and MyDIG. DAJ provided the microscopy images.  IF supervised the project. The manuscript
 was written by ATO, MFB and IF and was approved by all authors.

\section*{Funding}

This study was partially funded by National Institutes of Health [grant number R15
AI077394] awarded to MFB, and by National Science Foundation Advances in Bioinformatics
[grant number ABI 1146960] awarded to IF.  

\section*{Acknowledgements}

Will be added post-review.


{\ifthenelse{\boolean{publ}}{\footnotesize}{\small}
 \bibliographystyle{bmc_article}  
  \bibliography{biodig} }     


\ifthenelse{\boolean{publ}}{\end{multicols}}{}



\section*{Figures}
  \subsection*{Figure 1 - BioDIG Schema}
    BioDIG comprises three modules: Image, Genome and User Management. Data in the Genome Module
    is deposited initially by registered users, but gets updated via NCBI. Level 2 \& 3 users
    can upload images, and update image tags and image links to genomic data. 

  \subsection*{Figure 2 - Genome Module}

    The Genome Module consists of the GMOD PostGreSQL database, Chado which is used to house
    genome information. SQLite3 is used to interface with the GBrowse genome database, due to
    its speed. The duplication and maintenance of data integrity are not costly. See text for
    details.

  \subsection*{Figure 3 - User Management Module}
    User management is done by a circle of increasing privileges. Most users are Read
    Only, Level 1. Image and genome curators are Level 2 trusted users. Level 3 users can
    add and revoke user privileges, in additions to all Level 1 and 2 permissions.

  \subsection*{Figure 4 - Case study: \textit{Mycoplasma pneumoniae}}

    \textit{M. pneumoniae} image tagged with locations of attachment organelle gene products.
    \textit{M. pneumoniae} strain 19294 was passaged 10 times and grown and processed for
    imaging by scanning electron microscopy essentially as previously
    described\cite{pmid16804191}, but at higher magnification that is well-suited for
    morphological comparison across species. \\ \textbf{A}, (blue) image highlighting components of
    the proximal portion of the terminal organelle, consisting of the base of
    the electron-dense core. \textbf{B}, (green) image highlighting components
    of the long portion of the terminal organelle, consisting of the shaft, within which is
    the rod portion of the electron-dense core. Gene designations are keyed to the GenBank
    annotation of \textit{M. pneumoniae} strain M129.

\newpage
\includegraphics[width=0.7\textwidth]{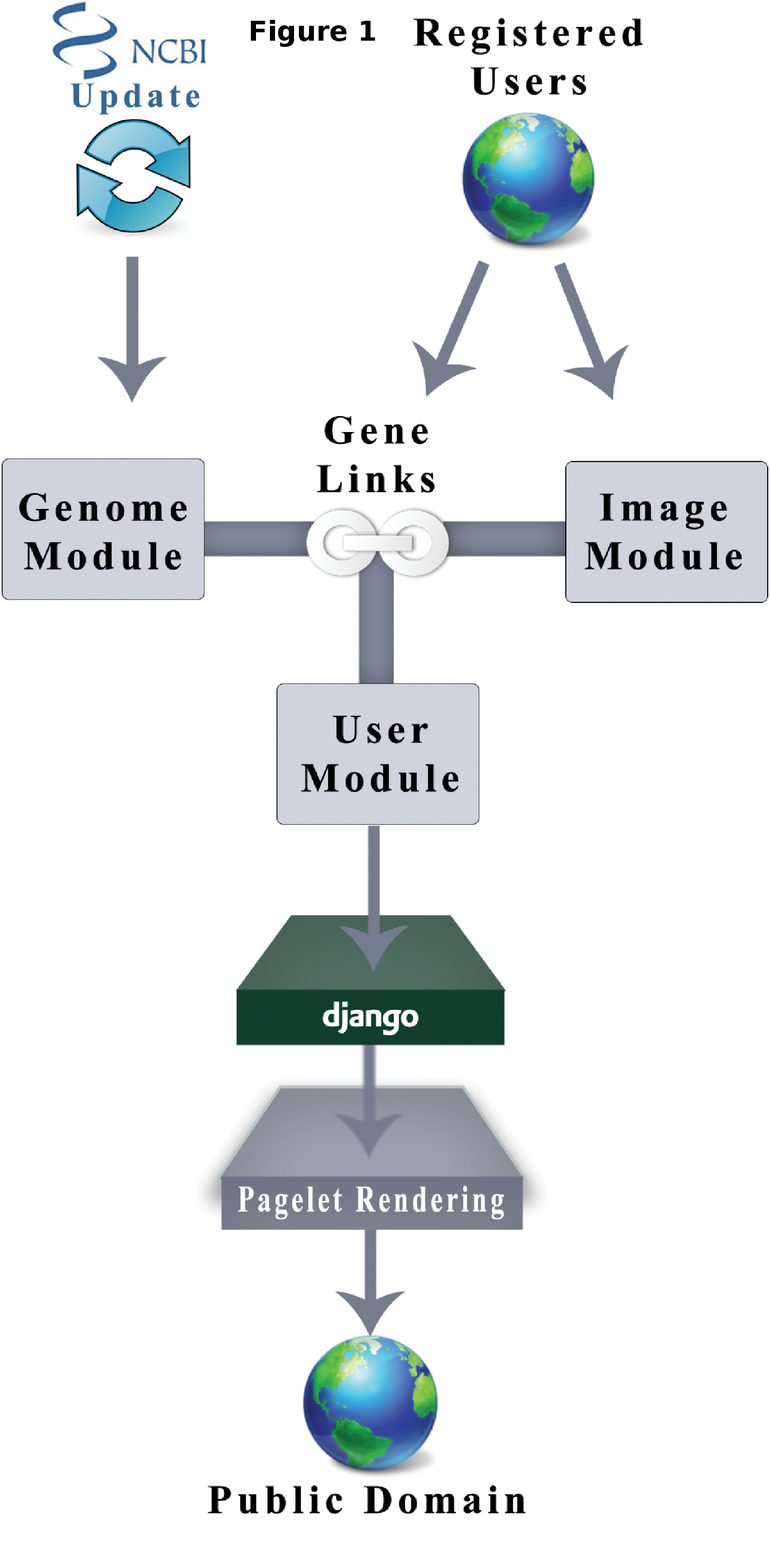}
\newpage
\includegraphics[width=1.0\textwidth]{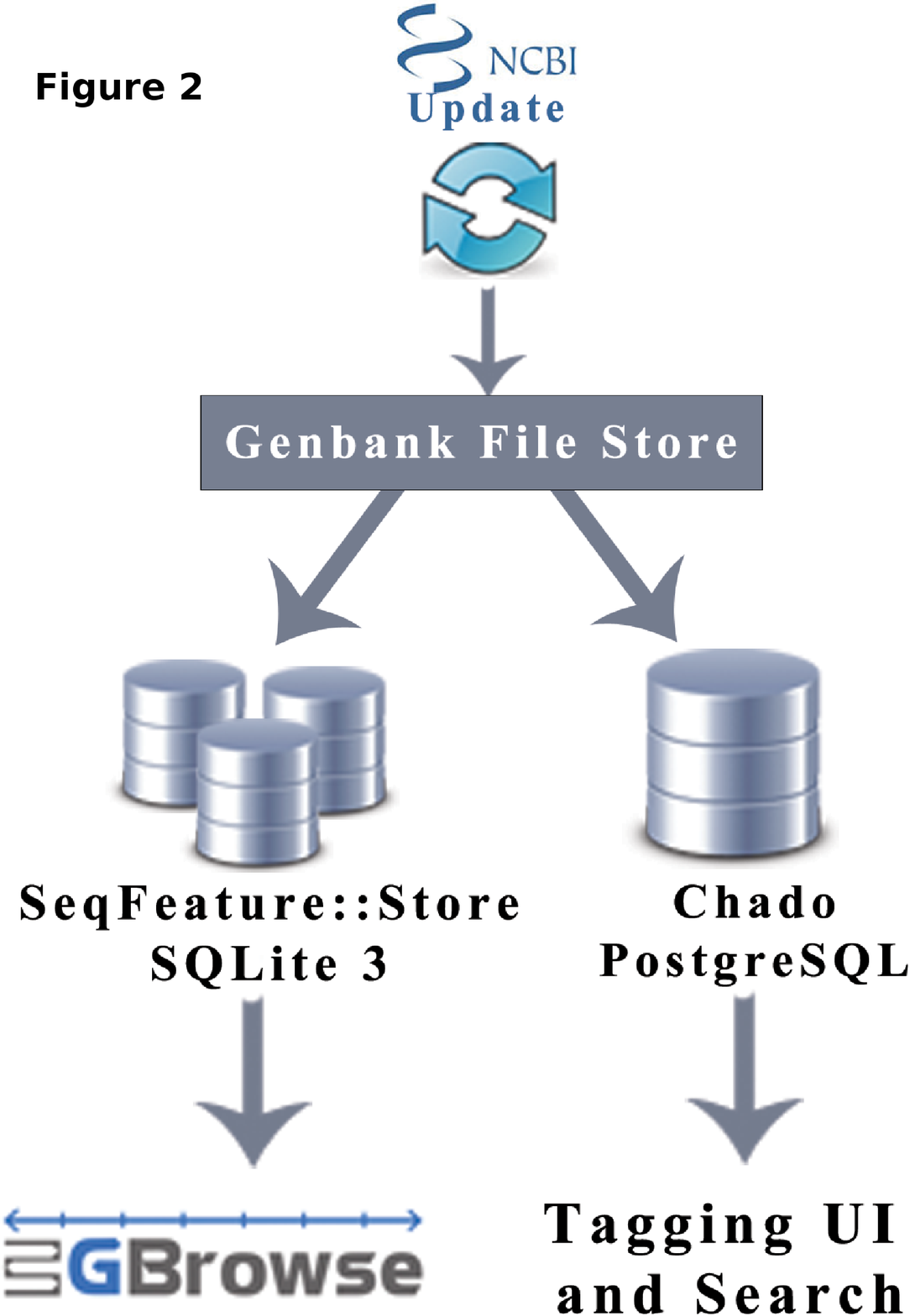}
\newpage
\includegraphics[width=1.0\textwidth]{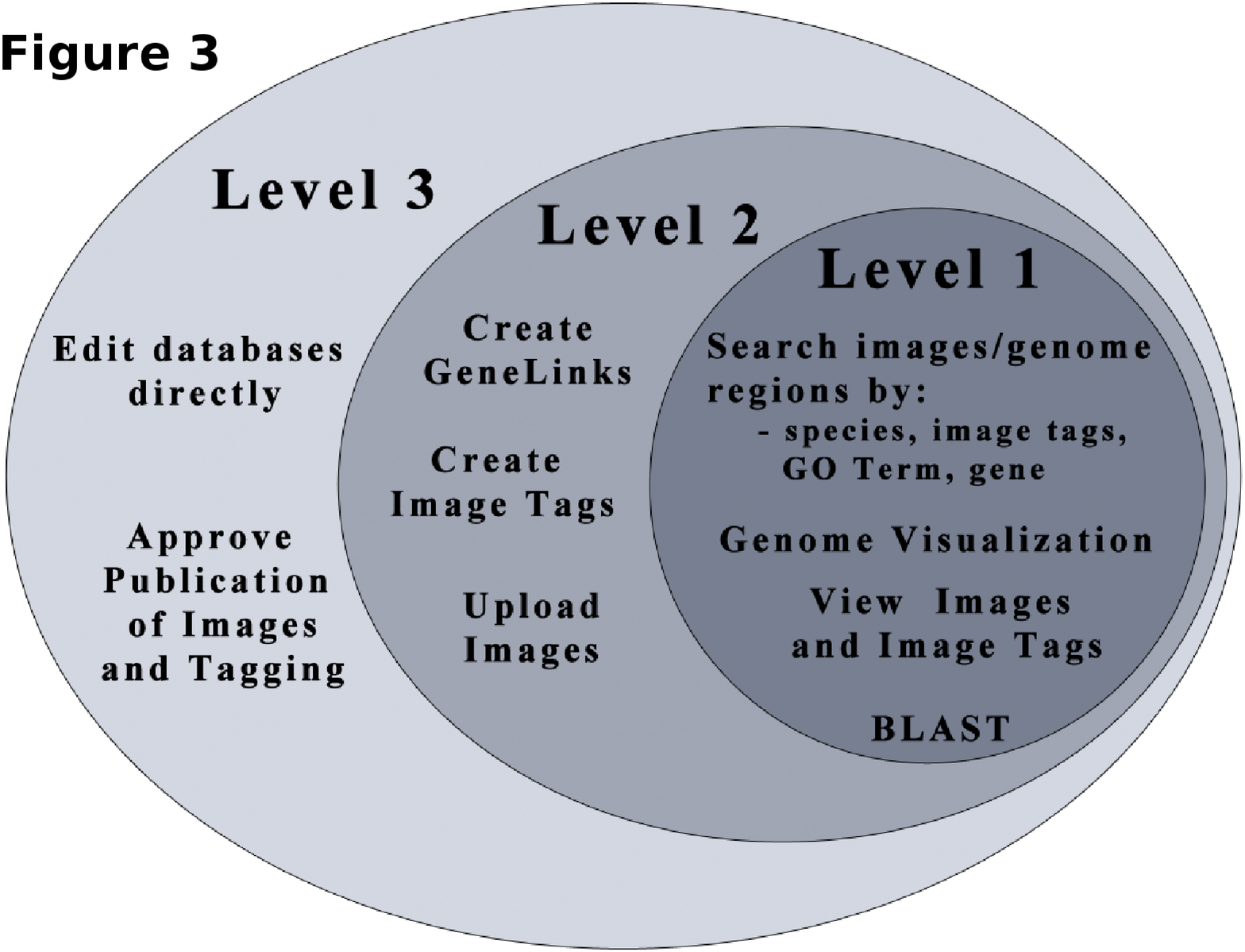}
\newpage
\includegraphics[width=1.0\textwidth]{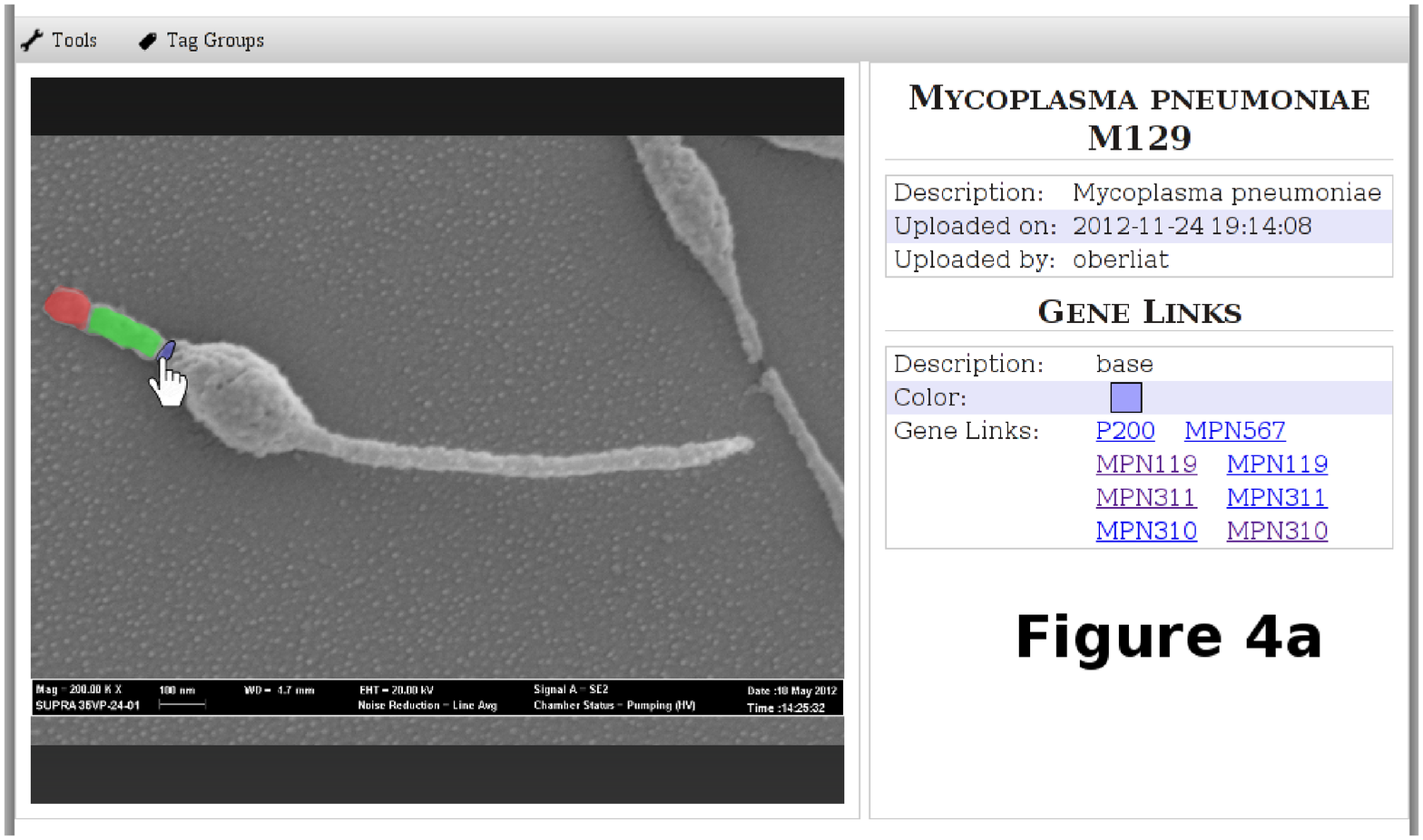}
\newpage
\includegraphics[width=1.0\textwidth]{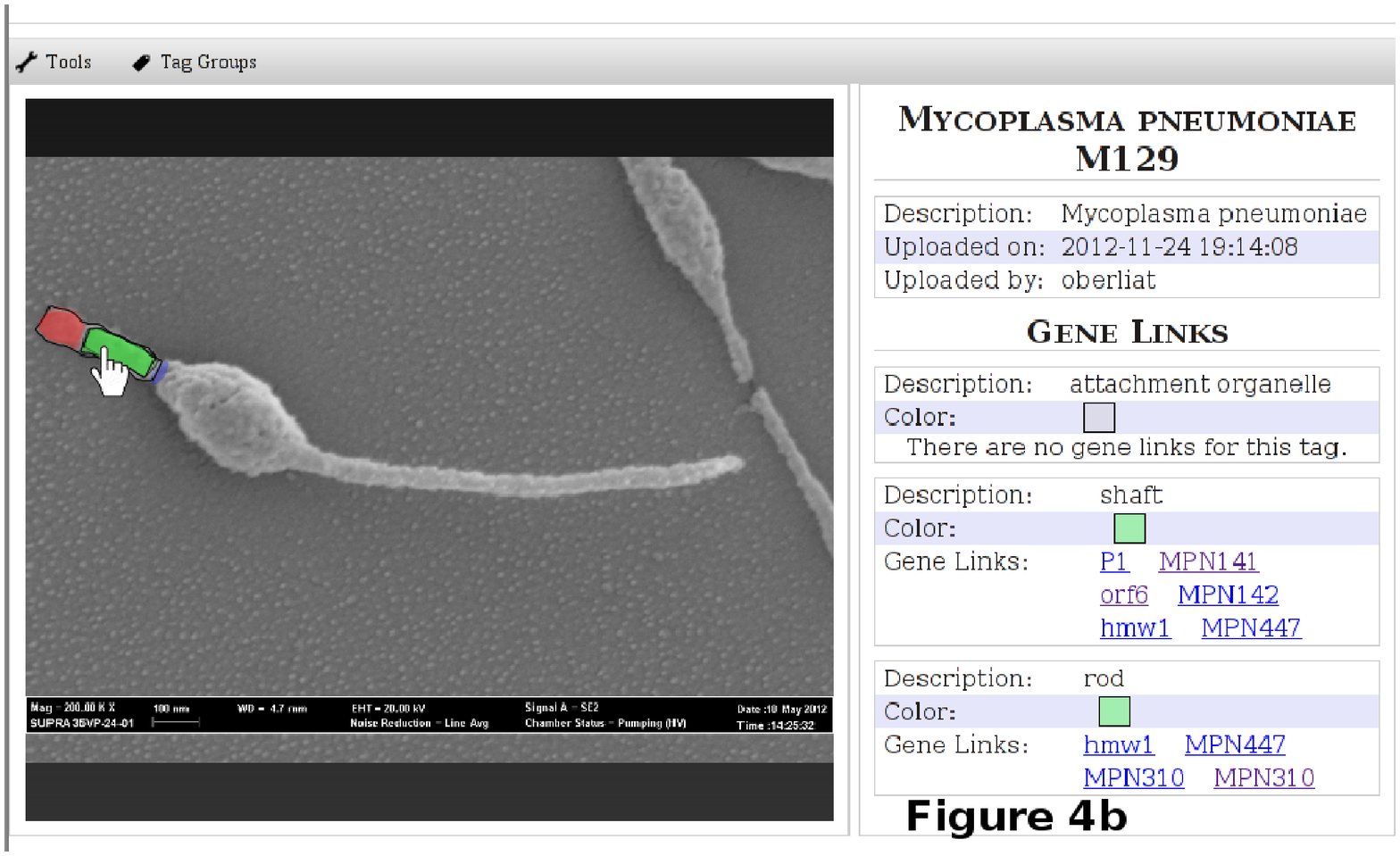}
\newpage




\end{bmcformat}
\end{document}